\documentclass[12pt]{article}
\usepackage[dvips]{graphicx}
\def\be{\begin{equation}}
\def\ee{\end{equation}}
\def\ba{\begin{eqnarray}}
\def\ea{\end{eqnarray}}

\begin{document}
\begin{titlepage}
\thispagestyle{empty}
\vskip0.5cm
\begin{flushright}
HUB-EP-98/26 \\
MS--TPI--98--6 
\end{flushright}
\vskip1.5cm

\begin{center}
{\Large {\bf 3D Ising Model with Improved}}
\vskip3mm

{\Large {\bf Scaling Behaviour}}
\end{center}

\vskip1.5cm
\begin{center}
{\large M. Hasenbusch${}^a$, K. Pinn${}^b$, and S. Vinti${}^b$}\\
\vskip5mm
${}^a$ Fachbereich Physik, Humboldt-Universit\"at zu Berlin\\ 
Invalidenstr.\ 110, D--10099 Berlin, Germany \\
e--mail: {\sl hasenbus@ficus1.physik.hu-berlin.de}
\vskip5mm
${}^b$ Institut f\"ur Theoretische Physik I, Universit\"at M\"unster\\ 
Wilhelm--Klemm--Str.~9, D--48149 M\"unster, Germany \\
e--mail: {\sl pinn@uni--muenster.de, vinti@uni--muenster.de}
\end{center}

\vskip2.5cm
\begin{abstract}
\par\noindent
We present results from the simulation of a two-coupling spin-1 model
with states $0,\pm 1$ and nearest neighbour interaction. By a suitable
choice of couplings we are able to drastically reduce the effects of
corrections to scaling. Our estimates for the critical exponents are
$\nu= 0.6299(3)$ and $\eta = 0.0359(10)$.  For the universal ratio $Q=
\langle m^2 \rangle^2 / \langle m^4 \rangle$ we obtain $Q= 0.6240(2)$.
The universal ratio of partition functions with antiperiodic/periodic
boundary conditions, respectively, is $Z_a/Z_p = 0.5425(2)$.
\end{abstract}
\end{titlepage}

\section{Introduction}
The determination of parameters like critical exponents 
from finite size scaling
\cite{privman}
is plagued by the appearance 
of corrections to scaling. Consider, e.g., an Ising model 
on a cubic lattice of linear extension $L$. 
The Binder cumulant  
\be
Q= \frac{\langle m^2 \rangle^2}{\langle m^4 \rangle} 
\ee 
behaves at the critical point like 
\be
\label{Qfss}
Q(L)= Q^* + q \, L^{-\omega} + \dots  
\ee
$m$ denotes the magnetization per spin. 
$Q^*$ is universal. 
$\omega \approx 0.8$ denotes the leading correction to scaling exponent. 
Many more terms appear in eq.~(\ref{Qfss}), governed by
subleading exponents and combinations of them. 
If they have non-negligible coefficients in front of them, 
they can hamper or even make impossible a reliable 
determination of the infinite $L$ behaviour.

In the language of the renormalization group the presence 
of strong corrections to scaling is related to the fact that 
one is far away from the renormalization group fixed point. 

In this article, we summarize the results of our attempt 
to improve the scaling properties of the 3D Ising model 
by suitably tuning the parameters in a two-coupling
spin-1 model that has already been studied by 
Bl\"ote et al.~\cite{bloete}.

We will be mainly interested in the determination of 
the exponents $\nu$ and $\eta$ that will be obtained 
from the scaling laws ($i=1,2$)
\be
\frac{\partial R_i}{\partial \beta}
= a_i \, L^{1/\nu} ( 1 + b_i \, L^{-\omega} + \dots)
\ee
and 
\be
\label{chifit1}
\chi = c + L^{2-\eta} ( 1 + d \, L^{-\omega} + \dots) \, .
\ee
Here, $R_2$ denotes the cumulant introduced above, and 
$R_1$ is the ratio of partition functions $Z_a/Z_p$, where
the subscripts label periodic and antiperiodic boundary 
conditions, respectively. Antiperiodic boundary conditions
are applied in one of the three lattice directions only. 
The susceptibility $\chi$ is defined by 
\be
\chi = L^3 \, \langle m^2 \rangle \, . 
\ee

\section{Improved Spin-1 Ising Model}

Following \cite{bloete}, we consider a spin-1 Ising model
on a simple cubic lattice defined through 
\be
S = - \beta \sum_{<i,j>} s_i \, s_j + D \sum_{i} s_i^2 \, . 
\ee 
The $s_i$ take values $0,\pm 1$, and the spin-spin interaction
is a sum over all nearest neighbour pairs. The Boltzmann factor is 
given by $\exp(-S)$.
Bl\"ote et al.\ chose a fixed $D=\ln 2$, and varied $\beta$
to tune to criticality. We followed a different procedure, 
to be described in some detail elsewhere.
    From prior simulations \cite{tocome}, estimates of 
the universal quantities $R_i^*$ were available, 
namely 
$R_1^* = 0.5425(10)$  and  $R_2^* = 0.624(1)$. 
For a number of lattice sizes $L_j$ ranging from 3 to 14  
a sequence of couplings $(\beta_j,D_j)$ was determined 
such that 
$R_i(\beta_j,D_j,L_j)= R_i^*$ within the statistical accuracy. 
As $L$ increases the pair $(\beta_j,D_j)$ converges to a point where
leading order corrections to scaling vanish. From our numerical results 
we extrapolate to
$(\beta_{\infty},D_{\infty})=(0.3832,0.6241)$.
We found that this point is approached on the line
\be
\label{critline}
  D \approx  -2.04 + 6.95 \, \beta \, .
\ee
It turns out that
eq.~(\ref{critline}) is an approximation to the critical line.  It is
reassuring that the estimate $\beta_c = 0.3934214(8)$, together with
$D= \ln 2$ in \cite{bloete} is close to this line.  Our next goal was
to approximately identify a second one-dimensional submanyfold of the
2-coupling space, namely that where the leading correction to scaling
vanishes. This was possible by fitting the derivatives of the
$R_i(\beta_j,D_j,L_j)$ 
at $(\beta_{\infty},D_{\infty})$
with respect to the two coupling parameters to
a certain linear equation. Details will be reported in \cite{tocome}.
Our fit results suggested that the critical line should be approached
by varying $\beta$ while adjusting $D$ according to 
\be
\label{goodline}
 D = 3 \, (\beta-0.3832) + 0.6241  \, .
\ee
A first estimate for $\beta_c$ was then obtained applying the well
established method looking for the crossings of  
$R_i(L)$ with $R_i(2\,L)$. We used lattices of size $L=4,8,16,32$
and obtained $\beta_c = 0.383245(10)$, with $D$ given by 
eq.~(\ref{goodline}), i.e.\ $D_c = 0.624235$. This is significantly
different from the $D= \ln 2$ used in ref.~\cite{bloete}.

\section{Simulation Results}

Monte Carlo simulations were now performed
at $\beta_c = 0.383245$, fixing $D$ according to 
eq.~(\ref{goodline}). 
We simulated on cubic lattices with linear extension 
$L$ ranging from 4 to 56. 
State of the art cluster algorithms were employed.  
Two measurements were separated by three single cluster updates.
For the computation of $Z_a/Z_p$ a variant of the   
boundary flip cluster algorithm \cite{flip}
was implemented. 
The runs for the determination of the critical parameters 
described in the previous section took about 3 months  
of CPU on Pentium 166-MMX PCs, while the final production runs consumed 
a total of 1 year.  

The data relevant for the present article are summarized 
in tables \ref{ZQX} and \ref{deri}. Table \ref{ZQX}
gives the quantities $R_i$, $i=1,2$ and $\chi/L^2$. 
In addition, we quote the number of measurements in
the last column. 
In table \ref{deri} we give the estimates of the 
partial and total derivatives of the $R_i$ with respect to 
$\beta$. Note that eq.~(\ref{goodline}) implies
$$
\frac{d R_i}{d \beta } = 
\frac{\partial R_i}{\partial \beta } 
+ 3 \, \frac{\partial R_i}{\partial D} \, .
$$
The quantities in table~\ref{deri} are multiplied by 
a factor $f(L)=L^{-1/0.63}$, which compensates to
good precision  for the leading divergent behaviour with
increasing lattice size. 

By a reweighting technique we have access also to the same set of
observables at a set of four neighbouring $\beta$ values, ranging from
0.383225 to 0.383265 in steps of 0.000010.  Having control over a
neighbourhood of the simulation point is important, since we started
without a high precision estimate of $\beta_c$.

\newcommand{\CC}{\phantom{1}}

\begin{table}
\begin{center}
\begin{tabular}{r|c|c|l|c}
$L$ & $Z_a/Z_p$ & $Q$ & \phantom{x} $\chi/L^2$ & stat/$10^6$ \\
\hline
 4& .53599(07)&.62408(05)& .87609(13) & 30 \\
 6& .54080(07)&.62491(06)& .88104(14) & 60 \\
 8& .54193(09)&.62460(07)& .87706(16) & 36 \\
10& .54209(18)&.62464(13)& .87272(33) & 10 \\
12& .54248(19)&.62416(14)& .86778(35) & 10 \\
14& .54230(21)&.62426(15)& .86399(38) & 10 \\
16& .54263(17)&.62409(13)& .86000(32) & 13  \\
18& .54251(23)&.62410(17)& .85697(42) &\CC  8 \\
20& .54286(28)&.62379(21)& .85332(50) &\CC  3 \\
22& .54244(30)&.62404(23)& .85115(55) &\CC  7\\
24& .54229(23)&.62402(17)& .84914(43) &\CC  3\\
28& .54245(31)&.62386(23)& .84373(55) &\CC  6\\
32& .54259(31)&.62390(23)& .84010(56) &\CC  6\\
36& .54213(41)&.62420(30)& .83742(73) &\CC  5\\
40& .54227(32)&.62403(23)& .83392(56) &\CC  7\\
48& .54250(39)&.62395(29)& .82803(69) &\CC  5\\
56& .54272(63)&.62382(47)& .82351(12) &\CC  2\\
\hline
 \end{tabular}
\parbox[t]{.85\textwidth}
 {
 \caption[ZQX]
 {\label{ZQX}
\small
Results for $R_1=Z_a/Z_p$, $R_2=Q$ and the susceptibility 
divided by $L^2$
at $\beta= 0.383245$. The last column gives the number of 
measurements in units of $10^6$. 
}
}
\end{center}
\end{table}

\begin{table}
\begin{center}
\begin{tabular}{r|l|l|l|l}
$L$ & $ f(L)\,\partial R_1/\partial\beta $ 
& $f(L)\, d R_1/d \beta$  
& $f(L)\, \partial R_2/\partial\beta$ 
& $ f(L)\, d R_2/d \beta$  \\
\hline
 4&--1.09974(17)&--.63210(11)& .65894(15)&--.09882(03)\\
 6&--1.13595(17)&--.64829(10)& .66236(15)&--.09685(03)\\
 8&--1.14621(20)&--.65290(12)& .66295(18)&--.09614(03)\\
10&--1.15051(41)&--.65498(24)& .66276(37)&--.09577(07)\\
12&--1.15293(45)&--.65567(25)& .66307(40)&--.09571(07)\\
14&--1.15385(49)&--.65661(27)& .66310(46)&--.09555(08)\\
16&--1.15432(44)&--.65652(25)& .66290(39)&--.09554(07)\\
18&--1.15547(59)&--.65692(33)& .66294(55)&--.09545(11)\\
20&--1.15603(71)&--.65732(39)& .66322(64)&--.09553(11)\\
22&--1.15542(79)&--.65714(43)& .66310(74)&--.09574(17)\\
24&--1.15675(61)&--.65790(34)& .66419(55)&--.09568(11)\\
28&--1.15565(80)&--.65730(44)& .66199(75)&--.09520(20)\\
32&--1.15664(82)&--.65793(46)& .66283(75)&--.09523(18)\\
36&--1.1568(12) &--.65776(61)& .6638(11) &--.09525(49)\\
40&--1.15665(84)&--.65785(47)& .66326(84)&--.09557(37)\\
48&--1.1563(11) &--.65752(59)& .6635(11) &--.09513(58)\\
56&--1.1562(17) &--.65739(97)& .6628(19) &--.0958(12)\\
\hline
 \end{tabular}
\parbox[t]{.85\textwidth}
 {
 \caption[deri]
 {\label{deri}
\small
Results for 
partial and total derivatives of 
$R_1=Z_a/Z_p$, $R_2=Q$, multiplied by the 
factor $f(L)=L^{-1/0.63}$, 
at $\beta= 0.383245$.
}
}
\end{center}
\end{table}

\section{Fitting the Data}

\subsection{Fitting $Z_a/Z_p$, $Q$}
We first fitted the $R_i$ in order to obtain estimates 
for the universal quantities $R_i^*$. The scaling law 
in question is eq.~(\ref{Qfss}). However, looking carefully
at table~\ref{ZQX} reveals that corrections to scaling 
are small. We therefore fitted the data with
the law 
\be
\label{fitR}
R_i(L,\beta_{MC})= R_i^* + \frac{d R_i}{d\beta}(L,\beta_{MC})
\, \Delta \beta \, . 
\ee
Here we have included a term which (to first order) corrects 
for deviations from being at criticality. $\beta_{MC}$ is 
our simulation coupling 0.383245, and the $d R_i/d\beta$
are taken from table~\ref{deri}.  
The fit parameters are $R_i^*$ and
$\beta_c$ entering through $\Delta \beta= \beta_{MC}-\beta_c$.
To check for effects from corrections to scaling, the fits were 
done on a sequence of data sets obtained by discarding data 
with $L < L_{\rm min}$. 
We first fitted separately $R_1$ and $R_2$. The results are 
reported in tables~\ref{RRR1} and \ref{RRR2}. Then we fitted all  
data together with three parameters ($R_1^*$, $R_2^*$, and $\beta_c$). 
The results are presented in table~\ref{RRRb}.

\begin{table}
\begin{center}
 \begin{tabular}{r|l|l|r}
$ L_{\rm min}$ &
\phantom{xx} $R_1^*$  &
\phantom{xxx} $\beta_c$ & 
  $\chi^2$/dof \\
 \hline
  6& 0.54143(5) & 0.3832516(8)  &  15.49 \\
  8& 0.54213(8) & 0.3832470(8)  &   1.78 \\
 10& 0.54240(10)& 0.3832453(9)  &   0.79 \\
 12& 0.54251(11)& 0.3832447(9)  &   0.49 \\
 16& 0.54260(15)& 0.3832442(11) &   0.46 \\
 20& 0.54252(25)& 0.3832446(14) &   0.55 \\
 24& 0.54226(37)& 0.3832457(20) &   0.27 \\
 \hline
 \end{tabular}
\parbox[t]{.85\textwidth}
 {
 \caption[RRR1]
 {\label{RRR1}
Fitting $R_1$ with eq.~(\ref{fitR})
}
}
\end{center}
\end{table}

\begin{table}
\begin{center}
 \begin{tabular}{r|l|l|r}
$ L_{\rm min}$ &
\phantom{xx} $R_2^*$  &
\phantom{xxx} $\beta_c$ & 
  $\chi^2$/dof \\
 \hline
  6& 0.62468(4) & 0.3832523(10)&  4.59 \\
  8& 0.62447(6) & 0.3832499(10)&  2.22 \\
 10& 0.62429(7) & 0.3832479(12)&  1.39 \\
 12& 0.62414(8) & 0.3832465(11)&  0.42 \\
 16& 0.62405(12)& 0.3832457(14)&  0.31 \\
 20& 0.62393(18)& 0.3832447(18)&  0.28  \\
 24& 0.62400(27)& 0.3832452(24)&  0.23  \\
 \hline
 \end{tabular}
\parbox[t]{.85\textwidth}
 {
 \caption[RRR2]
 {\label{RRR2}
Fitting $R_2$ with eq.~(\ref{fitR})
}
}
\end{center}
\end{table}

\begin{table}
\begin{center}
 \begin{tabular}{r|l|l|l|r}
$ L_{\rm min}$ &
\phantom{xx} $R_2^*$  &
\phantom{xx} $R_1^*$  &
\phantom{xxx} $\beta_c$ & 
  $\chi^2$/dof \\
 \hline
 6& 0.62467(4) & 0.54142(4) & 0.3832519(5) & 10.05 \\
 8& 0.62441(5) & 0.54206(7) & 0.3832481(6) &  2.19 \\
10& 0.62421(6) & 0.54231(9) & 0.3832463(7) &  1.21 \\
12& 0.62408(7) & 0.54245(11)& 0.3832453(8) &  0.51 \\
16& 0.62399(8) & 0.54254(14)& 0.3832448(9) &  0.42 \\
20& 0.62393(13)& 0.54252(22)& 0.3832446(11)&  0.41 \\
24& 0.62403(13)& 0.54230(22)& 0.3832455(11)&  0.25 \\
 \hline
 \end{tabular}
\parbox[t]{.85\textwidth}
 {
 \caption[RRRb]
 {\label{RRRb}
Fitting simultaneously $R_2$ and $R_1$ with eq.~(\ref{fitR})
}
}
\end{center}
\end{table}

\begin{figure}
\begin{center}
\includegraphics[width=11cm]{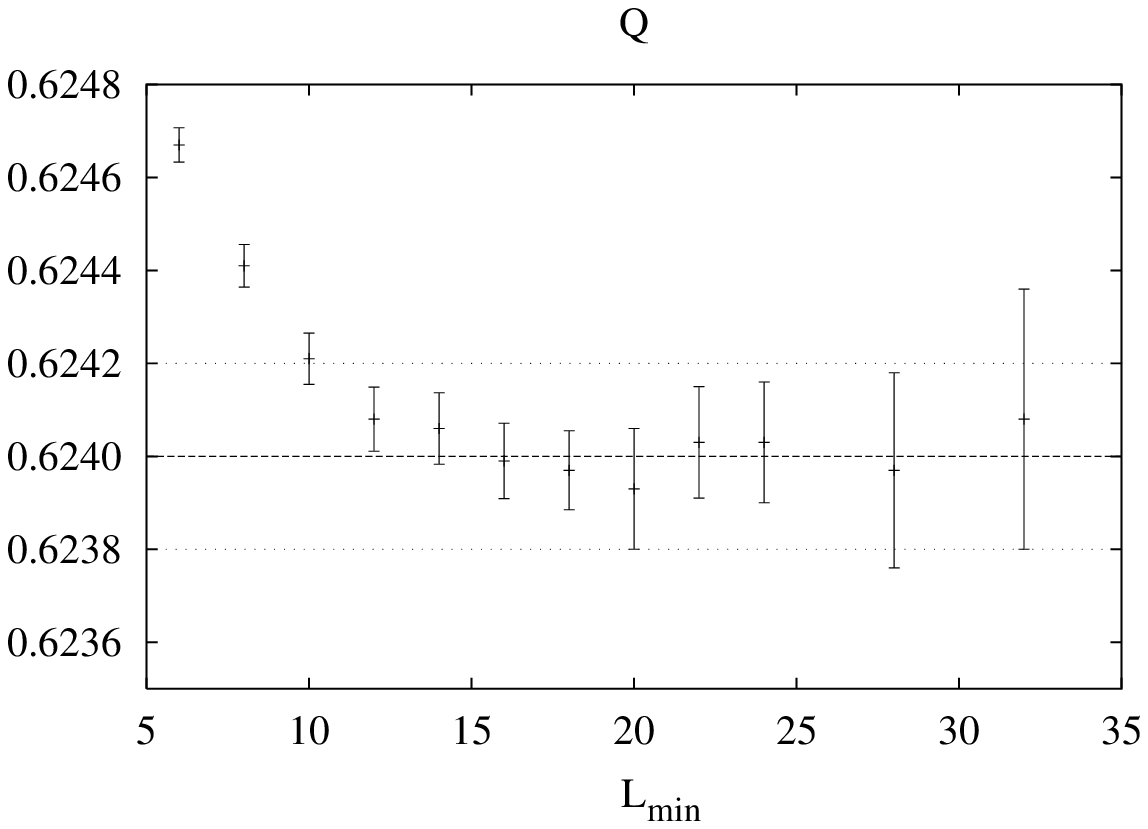}
\parbox[t]{.85\textwidth}
 {
 \caption[figQ]
 {\label{figQ}
 Fit results for $R_2^*$ as function of smallest lattice 
included in the fit.
 }
 }
\end{center}
\end{figure}

\begin{figure}
\begin{center}
\includegraphics[width=11cm]{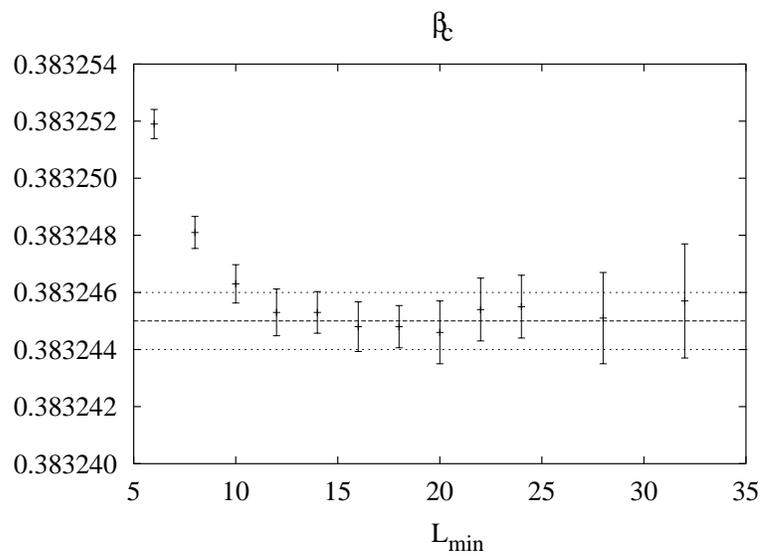}
\parbox[t]{.85\textwidth}
 {
 \caption[figB]
 {\label{figB}
\small
 Fit results for $\beta_c$ as function of smallest lattice 
included in the fit.
 }
 }
\end{center}
\end{figure}

Our fit results for $Q^*$ and the $\beta_c$'s from the $Q$-fits 
(as function of $L_{\rm min}$) are shown in 
figures~\ref{figQ} and \ref{figB}. The plots justify 
our final estimates $R_2^*= 0.6240(2)$ and 
$\beta_c= 0.383245(1)$. For $R_1^*$ we find 
0.5425(2). 

\subsection{Fitting the Derivatives of $R_i$}

\begin{figure}
\begin{center}
\includegraphics[width=11cm]{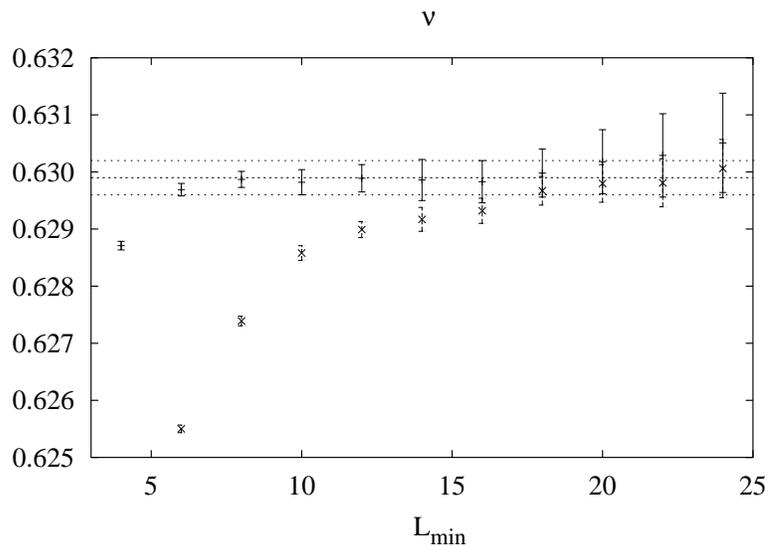}
\parbox[t]{.85\textwidth}
 {
 \caption[zqder]
 {\label{zqder}
Fit results for $\nu$ from fitting $\partial R_i/\partial \beta$ 
with eq.~(\ref{fitQp}). The data points with the better scaling
behaviour belong to the cumulant $R_2$.
 }
 }
\end{center}
\end{figure}

We fitted our results for the derivatives of the $R_i$ with 
respect to $\beta$, according to the law
\be
\label{fitQp}
\frac{ \partial R_i}{\partial \beta} = a_i \, L^{1/\nu} \, .  
\ee
Our results for $a_2$, i.e., for the derivative of the 
cumulant, are presented in table~\ref{Qp}. The $\nu$-results of
this table, together with the corresponding results for 
$Z_a/Z_p$ are plotted in figure~\ref{zqder}. Obviously, the 
derivatives of the cumulant scale much better than those 
of $Z_a/Z_p$.

For the derivative of $R_1$,
including the
leading correction to scaling {\em plus} a term $\sim L^{-x}$ with $x$
of order 2 improved the fits.  Note, however, that there is more than
one exponent or combination of exponents of order 2 (e.g,. $1/\nu+\omega$
or $\omega'$), and fits of this type do not have a sound theoretical basis.
The nice scaling behaviour of our data (especially of the cumulant
related quantities) allows us to avoid getting into
the mess of multi-parameter fits.

\begin{figure}
\begin{center}
\includegraphics[width=11cm]{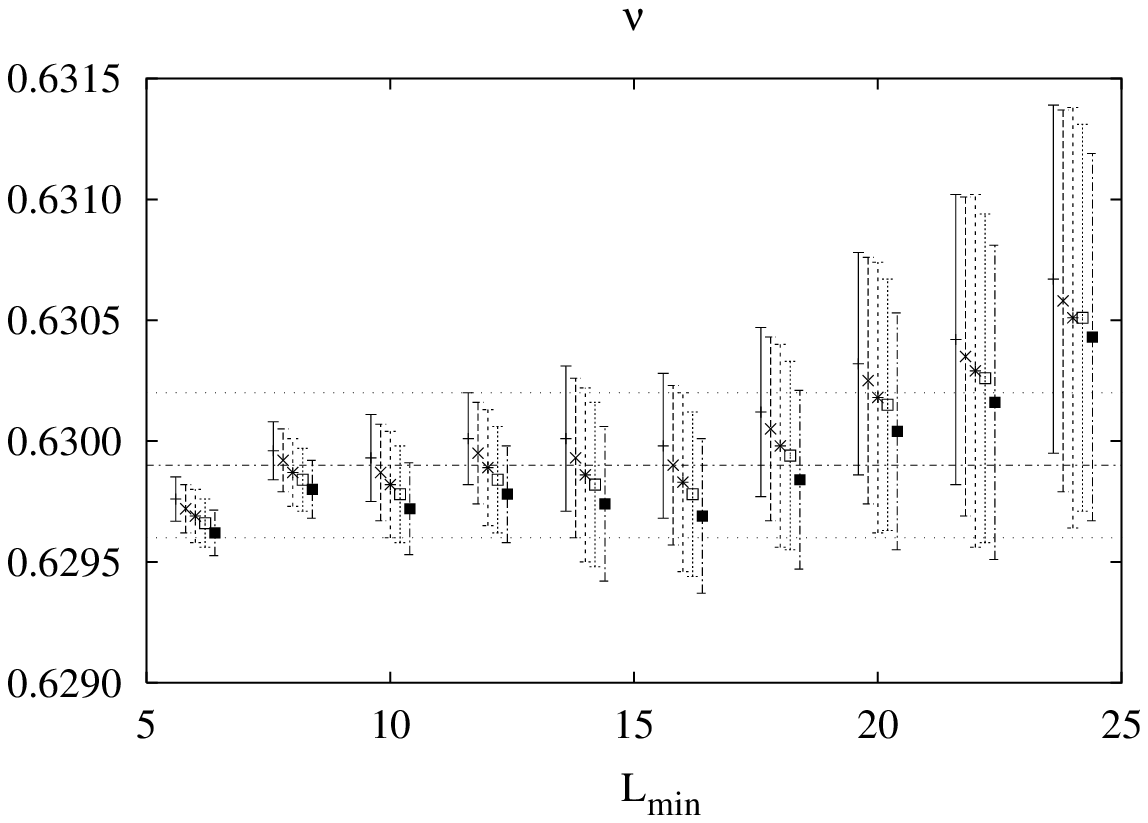}
\parbox[t]{.85\textwidth}
 {
 \caption[betanu]
 {\label{betanu}
Fit results for $\nu$ from fitting $\partial R_2/\partial \beta$ 
with eq.~(\ref{fitQp}). To check for effects of errors in the 
location of $\beta_c$, 
the $\nu$-estimate is given for 
the five $\beta$-values 0.383243 to 0.383247 in steps of 0.000001
(left to right).
 }
 }
\end{center}
\end{figure}

To check for the systematic dependence on the location of 
$\beta_c$ we repeated the fit for the $Q$-derivative on 
data  from five 
shifted $\beta$-values from 
0.383243 to 0.383247 in steps of 0.000001, covering thus 
two standard deviations around our $\beta_c$ estimate. 
Figure~\ref{betanu} shows that the variation of the 
$\nu$ estimate to this $\beta$ allows us to put as 
our final estimate 
\be 
\nu= 0.6299(3).
\ee

\begin{table}
\begin{center}
 \begin{tabular}{r|l|l|r}
$ L_{\rm min}$ &
\phantom{xx} $a_2$  &
\phantom{xx} $\nu$  &
  $\chi^2$/dof \\
 \hline
 6& 0.66160(42)& 0.62969(11) &  0.79 \\
 8& 0.66247(63)& 0.62987(14) &  0.57 \\
10& 0.6622(11) & 0.62982(22) &  0.60 \\
12& 0.6626(12) & 0.62989(24) &  0.64 \\
14& 0.6624(18) & 0.62986(36) &  0.70 \\
16& 0.6622(19) & 0.62983(37) &  0.77 \\
18& 0.6631(23) & 0.62998(42) &  0.84 \\
20& 0.6643(31) & 0.63018(56) &  0.90 \\
24& 0.6662(51) & 0.63051(87) &  1.18 \\
 \hline
 \end{tabular}
\parbox[t]{.85\textwidth}
 {
 \caption[Qp]
 {\label{Qp}
Fitting the derivative of $R_2$ with respect to $\beta$. The fit
function is given in eq.~(\ref{fitQp}).
}
}
\end{center}
\end{table}

\subsection{Fitting the Susceptibility}

\begin{figure}
\begin{center}
{\includegraphics[height=7cm]{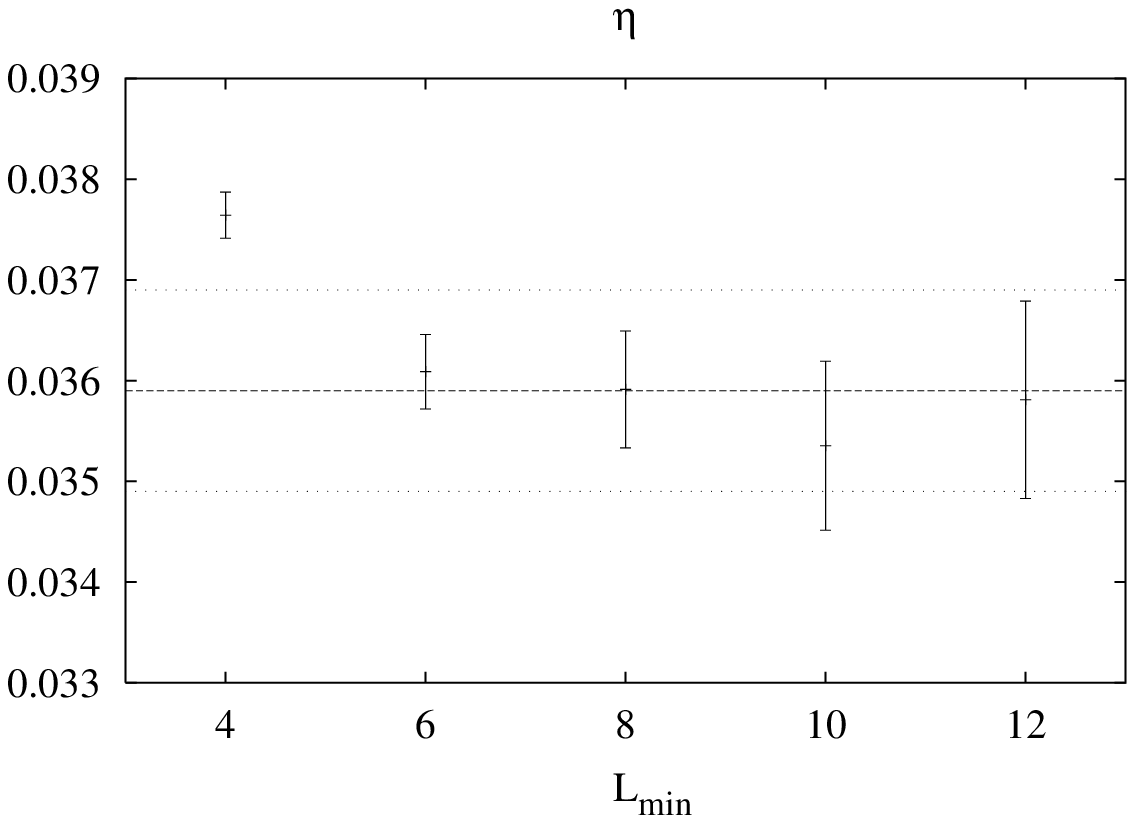}}
{\includegraphics[height=7cm]{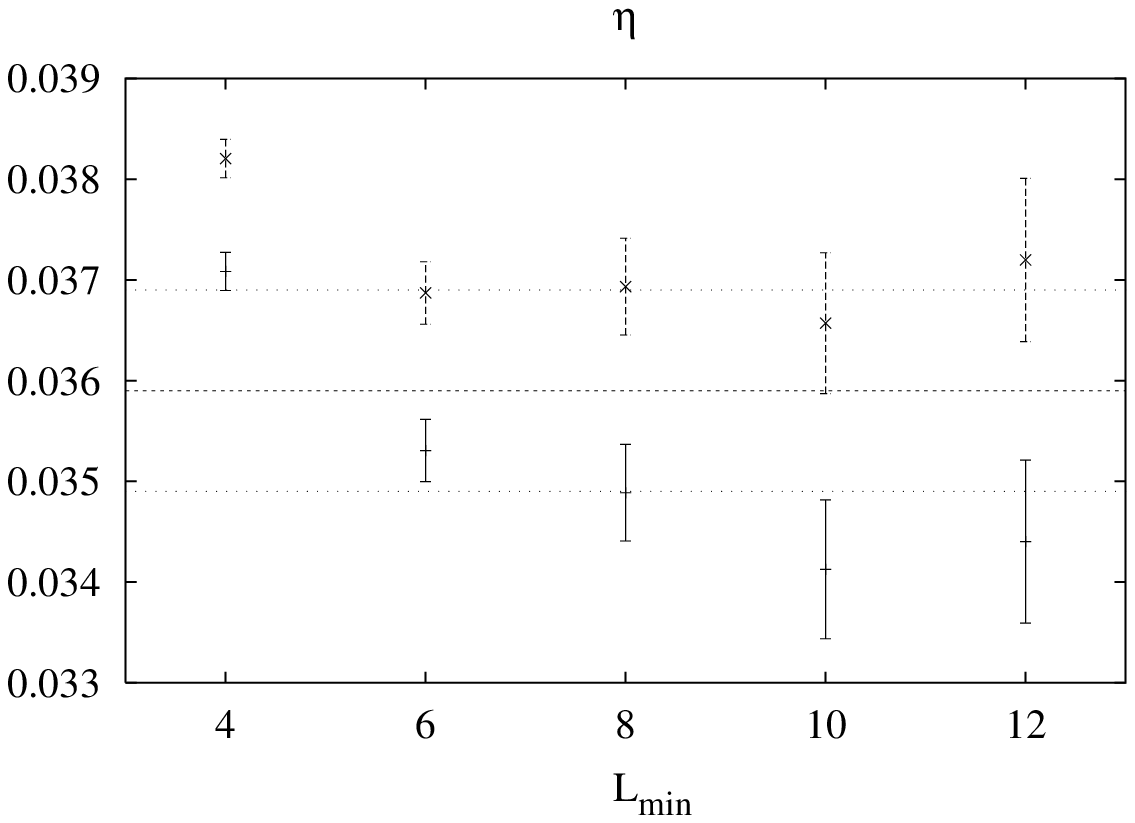}}
\parbox[t]{.85\textwidth}
 {
 \caption[aa2]
 {\label{aa2}
$\eta$ estimates
from fit function eq.~(\ref{chifit1}) 
as function of $L_{\rm min}$. Top: at our estimated 
value for $\beta_c= 0.383245$. Bottom: at $\beta$ 
chosen two standard deviations above (upper data) and below
(lower data) our $\beta_c$ estimate, i.e.\ at $\beta=0.383243$ 
and $\beta= 0.383247$.
 }
 }
\end{center}
\end{figure}

We fitted the susceptibility data given in table~\ref{ZQX}, and the
corresponding data at shifted $\beta$-values with eq.~(\ref{chifit1}).
It turned out that the first correction to scaling is negligible.
However, the fit results depend to some extent on the location of
$\beta_c$. The results for the exponent $\eta$ for $\beta_c =
0.383245$ are given in table~\ref{ccc1}.  Figure~\ref{aa2} shows the
$\eta$ estimates from table~\ref{ccc1} and also those from
$\beta$-values chosen two standard deviations above (upper data) and
below (lower data) our $\beta_c$ estimate.

\begin{figure}
\begin{center}
{\includegraphics[width=12cm]{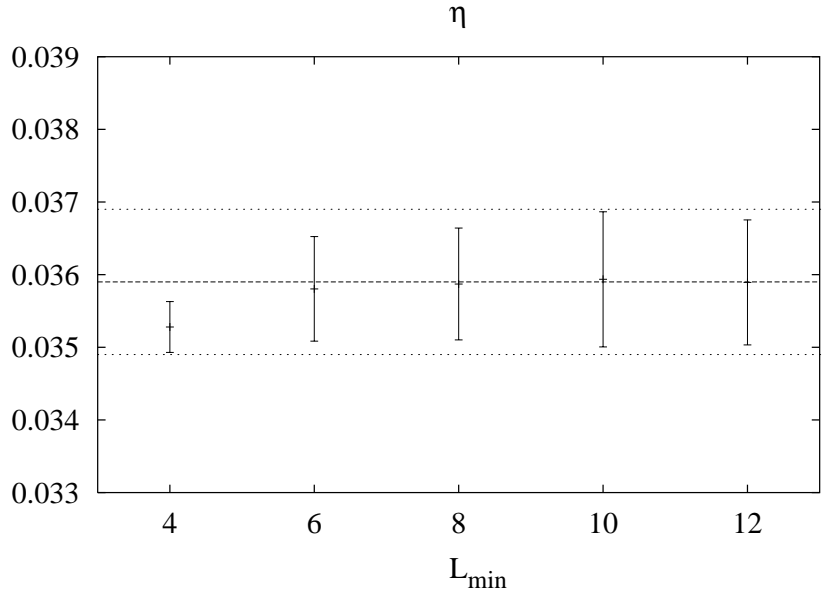}}
\parbox[t]{.85\textwidth}
 {
 \caption[aa1]
 {\label{aa1}
$\eta$ estimates as function of $L_{\rm min}$, fitting 
$\chi$-data at fixed $Q$ to eq.~(\ref{fixi}).
 }
 }
\end{center}
\end{figure}

We then tried to do the fits in an alternative way. 
Define a function $\beta_c(L)$ by requiring that 
for any $L$ the relation $Q^*(L,\beta_c(L))= Q^* = 0.6240$. 
The susceptibility as a function of $L$ and $\beta_c(L)$ 
should behave to leading order as 
\be\label{fixi}
\chi~(L, \beta_c(L)) = \tilde c +  \tilde d \,  L^{2-\eta} \, . 
\ee
Our fit results are given in table~\ref{ccc2} and plotted 
in figure~\ref{aa1}.
As a final estimate for $\eta$ we quote 
\be
\eta = 0.0359(10) \, . 
\ee

\begin{table}
\begin{center}
 \begin{tabular}{r|l|l|c|c}
$ L_{min}$ 
& \phantom{xxx}$c$ & \phantom{xxx} $d$ & $\eta$ 
&  $\chi^2$/dof \\
 \hline
  4& --0.5216(56)& 0.95751(59)& 0.03764(23) &  2.30 \\
  6& --0.429(17) & 0.9526(11) & 0.03609(37) &  0.51 \\
  8& --0.412(47) & 0.9520(19) & 0.03591(58) &  0.54 \\
 10& --0.32(11)  & 0.9501(28) & 0.03535(84) &  0.49 \\
 12& --0.42(17)  & 0.9517(34) & 0.03581(98) &  0.50 \\
 \hline
 \end{tabular}
\parbox[t]{.85\textwidth}
 {
 \caption[ccc1]
 {\label{ccc1}
Results from fitting the susceptibilities at $\beta=0.383245$
to eq.~(\ref{chifit1}).
}
}
\end{center}
\end{table}

\begin{table}
\begin{center}
 \begin{tabular}{r|l|l|c|c}
$ L_{min}$ 
& \phantom{xxx}$\tilde c$ & \phantom{xxx}$\tilde d$ & $\eta$ 
&  $\chi^2$/dof \\
 \hline
 4 &--0.4631(95)& 0.9502(10)& 0.03528(35) & 0.27 \\
 6 &--0.551(93) & 0.9520(24)& 0.03580(72) & 0.22 \\
 8 &--0.56(12)  & 0.9522(26)& 0.03587(77) & 0.24 \\
 10&--0.58(15)  & 0.9524(31)& 0.03594(93) & 0.26 \\
 12&--0.57(14)  & 0.9523(29)& 0.03589(86) & 0.28 \\
\hline
\end{tabular}
\parbox[t]{.85\textwidth}
{
\caption[ccc2]
{\label{ccc2}
Fitting the susceptibility at fixed $Q=0.6240$. 
}
}
\end{center}
\end{table}

\begin{table}
\begin{center}
 \begin{tabular}{c|l|l|l}
Ref. & Method & \phantom{xxx}$\nu$ &\phantom{xx} $\eta$ \\
 \hline
\cite{nickel}   & 3D FT & 0.6301(5)  & 0.0355(9)  \\   
\cite{guida}    & 3D FT & 0.6304(13) & 0.0335(25) \\  
\cite{guida}    & EPS   & 0.6293(26) & 0.036(6)   \\   
\cite{butera}   & HT    & 0.6310(5)  &            \\   
\cite{bloete}   & MC    & 0.6301(8)  & 0.037(3)   \\   
\cite{hapi}     & MC    & 0.6308(10) &            \\
\hline 
present work    & MC    & 0.6299(3)  & 0.0359(10) \\ 
\hline
\end{tabular}
\parbox[t]{.85\textwidth}
{
\caption[exp1]
{\label{exp1}
Comparing the results of the present study with previous
estimates in the literature. 3D FT means field theoretic calculations 
directly in three dimensions, EPS refers to $\epsilon$ expansion, and
MC means that the estimate relies on Monte Carlo simulations. 
}
}
\end{center}
\end{table}

\section*{Conclusions}
We have demonstrated that  
the spin-1 Ising model with suitably chosen coupling 
constants has remarkably improved finite size scaling
properties. This allowed us to obtain high precision estimates for the 
critical indices $\nu$ and $\eta$ and two other universal quantities, 
the Binder cumulant $Q$ and the ratio of partition functions
$Z_a/Z_p$.

Our results compare very well with a selection of estimates obtained 
by other authors, quoted in table~\ref{exp1}. Many more estimates 
can be found in refs.~\cite{bloete} and \cite{guida}.
Our estimate for the cumulant, $Q^*= 0.6240(2)$ may be compared with
the one given in \cite{bloete}, namely 
$Q^*= 0.6233(4)$. 

It would be worthwhile to use the present model in studies of physical
quantities not discussed in this work and to check to what extent the
improved scaling behaviour helps to get better estimates.

Of course, one could also search for even further reduction of
corrections to scaling by refining our procedure or choosing variants
of the model, e.g., the $\phi^4$-model or an Ising model with more
couplings. 
Last but not least, application of the ideas underlying the 
present analysis to other models seems promising.


\begin{thebibliography}{99}

\bibitem{privman} V. Privman, in: Finite Size Scaling and Numerical
                  Simulations of Statistical Systems, V. Privman, 
                  ed., World Scientific, Singapore, 1990. 

\bibitem{bloete} H.W.J. Bl\"ote, E. Luijten, and J.R. Heringa,
                 J.~Phys.~A 28 (1995) 6289, cond-mat/9509016. 

\bibitem{tocome} M. Hasenbusch, K. Pinn, and S. Vinti, in preparation.

\bibitem{flip}   M. Hasenbusch, Physica~A 197 (1993) 423.  

\bibitem{guida}  R. Guida and J. Zinn-Justin, cond-mat/9803240. 

\bibitem{nickel} Results by Murray and Nickel, taken from table 10 
                 of \cite{guida}. Errors from uncertainty of 
                 $\tilde g^*$ are not taken into account.

\bibitem{butera} P. Butera and M. Comi, Phys.\ Rev.\ B 56 (1997) 8212, 
                 hep-lat/9703018. An estimate for $\eta$ may be obtained 
                 using $\gamma = 1.2385(5)$ and the relation 
                 $\eta = 2 - \gamma/\nu$. 

\bibitem{hapi}   M. Hasenbusch and K. Pinn, cond-mat/9706003, to appear
                 in J.\ Phys.\ A. In this work, also $\alpha= 0.1115(37)$
                 is obtained. 

\end{thebibliography}
\end{document}